\documentclass[12pt]{article}

\usepackage[margin=1in]{geometry}
\usepackage{setspace}
\usepackage{lineno}

\usepackage{amsmath,amssymb,amsthm,bm,mathtools}
\usepackage{graphicx}
\usepackage{booktabs,multirow,threeparttable}
\usepackage{float,rotating}
\usepackage{siunitx}
\usepackage{array}
\usepackage{url}
\usepackage{xcolor}
\usepackage{hyperref}
\usepackage{natbib}
\defcitealias{early2015aromatase}{EBCTCG, 2015}
\defcitealias{breast2005comparison}{BIG 1-98, 2005}

\usepackage{cancel}
\usepackage{bbm}
\usepackage{caption}
\usepackage{soul}

\usepackage{calc}

\theoremstyle{plain}

\theoremstyle{definition}

\theoremstyle{remark}

\title{Locally Interpretable Individualized Treatment Rules for Black-Box Decision Models}

\author{
  Yasin Khadem Charvadeh\textsuperscript{1}, Katherine S. Panageas\textsuperscript{1}, and Yuan Chen\textsuperscript{1}\\[0.5em]
  \small \textsuperscript{1}Department of Epidemiology \& Biostatistics,
  \small Memorial Sloan Kettering Cancer Center,\\
  \small New York, New York, USA
}
\date{}
\begin{document}

\maketitle

\begin{abstract}
Individualized treatment rules (ITRs) aim to optimize healthcare by tailoring treatment decisions to patient-specific characteristics. Existing methods typically rely on either interpretable but inflexible models or highly flexible black-box approaches that sacrifice interpretability; moreover, most impose a single global decision rule across patients. We introduce the Locally Interpretable Individualized Treatment Rule (LI-ITR) method, which combines flexible machine learning models to accurately learn complex treatment outcomes with locally interpretable approximations to construct subject-specific treatment rules. LI-ITR employs variational autoencoders to generate realistic local synthetic samples and learns individualized decision rules through a mixture of interpretable experts. Simulation studies show that LI-ITR accurately recovers true subject-specific local coefficients and optimal treatment strategies. An application to precision side-effect management in breast cancer illustrates the necessity of flexible predictive modeling and highlights the practical utility of LI-ITR in estimating optimal treatment rules while providing transparent, clinically interpretable explanations.
\end{abstract}

\newpage

\section{Introduction}\label{intro}
Precision medicine represents a transformative approach to healthcare aimed at preventing, diagnosing, treating, and monitoring diseases through highly individualized healthcare strategies \citep{lesko2007personalized}. Unlike traditional approaches that often rely on a ``one-size-fits-all'' approach to care, precision medicine tailors healthcare interventions to the unique characteristics of each patient. This personalized approach can integrate a wide array of patient-specific factors, including genetic profiles, molecular biomarkers, phenotypic traits, and even psychosocial characteristics \citep{jameson2015precision}. By accounting for the interindividual variability in treatment response, precision medicine has the potential to improve both the effectiveness and safety of therapeutic interventions and care delivery strategies. This approach is particularly impactful in the management of chronic diseases, such as cancer, where long-term treatment plans must be continuously adapted to meet the evolving needs of patients. By emphasizing individualized care, precision medicine holds the promise of improving clinical outcomes, reducing adverse effects, and advancing overall public health.

Within this framework, a treatment that is customized to the specific characteristics of an individual patient in a single-stage study design is referred to as an \textit{individualized treatment rule} (ITR) \citep{murphy2001marginal}. 
Developing optimal ITRs typically involves either \textit{indirect} or \textit{direct} methodological approaches \citep{laber2014dynamic}. Indirect methods, such as Q-learning \citep{watkins1989learning}, A-learning \citep{murphy2003optimal}, and G-estimation \citep{robins2004optimal}, first model the conditional mean outcome and subsequently derive the optimal ITRs. These approaches allow for counterfactual reasoning, enabling predictions about how different treatment choices might alter outcomes. However, they are often sensitive to model misspecification, which can lead to biased or inaccurate treatment rules \citep{laber2014dynamic}. While more flexible models can improve the estimation of conditional mean outcomes, this often comes at the expense of the interpretability of the estimated treatment rule, thereby limiting its clinical applicability as a decision support tool \citep{kosorok2019precision}. In contrast, direct methods, including outcome-weighted learning (OWL) \citep{zhao2012estimating} and marginal structural mean models \citep{robins2008estimation}, focus on optimizing treatment decisions without explicitly modeling the conditional mean outcome. These methods improve robustness against model misspecification, resulting in more reliable treatment rules under varying conditions. However, this robustness comes with a trade-off; namely, higher variance in the resulting estimators \citep{laber2014dynamic}, which can affect the stability of the treatment recommendations.

The choice of an appropriate method for developing ITRs hinges on balancing interpretability, robustness, and computational feasibility. Traditional regression-based models are highly interpretable and transparent, making them well-suited for clinical decision-making, but their ability to capture complex, nonlinear effects is limited. In contrast, black-box machine learning (ML)-based approaches offer greater flexibility for modeling nonlinear effects \citep{mi2019bagging, zhao2012estimating}, often at the cost of interpretability. Striking an appropriate balance among these priorities is crucial for developing ITRs that are not only statistically rigorous but also clinically meaningful, practical, and implementable in real-world settings.

Furthermore, most existing approaches to developing ITRs focus on deriving a single, universal treatment rule that is applied uniformly to all patients. While the resulting treatment recommendations may be tailored to individual characteristics, the underlying decision rule itself remains essentially one-size-fits-all. Such global rules may fail to capture the subtle, patient-specific heterogeneity in treatment effects. For example, the relationship between CYP2D6 genotype, tamoxifen therapy, and breast cancer prognosis illustrates this clinical complexity that can arise when treatment effects vary across individuals. \cite{he2020cyp2d6} demonstrated that CYP2D6 metabolizer status is a critical factor in determining the effectiveness of tamoxifen in treating estrogen-receptor-positive breast cancer. Poor metabolizers struggle to convert tamoxifen into its active form, endoxifen, leading to a higher risk of breast cancer-specific mortality with a hazard ratio (HR) of 2.59. On the other hand, ultrarapid metabolizers also face worse outcomes (HR = 4.52), possibly due to excessive drug metabolism causing suboptimal endoxifen levels or increased side effects, which may contribute to higher treatment discontinuation rates (18.8\% vs. 6.7\% in normal metabolizers). Their study found a U-shaped relationship between CYP2D6 metabolizer status and breast cancer-specific mortality in patients treated with tamoxifen. Therefore, the treatment benefit may vary in diverse and even opposite ways depending on patients' own CYP2D6 metabolizer status. This highlights the need for more refined, genotype-informed treatment rules that can account for such complex and multifaceted treatment mechanisms.
In the ML literature, several methods have been developed to improve the interpretability of complex black-box models, which are well-suited to capture intricate treatment mechanisms. One widely adopted approach is Local Interpretable Model-agnostic Explanations (LIME) \citep{ribeiro2016should}, which explains model predictions by attributing them to individual input features, thereby increasing transparency in the decision-making process. LIME works by approximating the behavior of a black-box model within localized regions of the input space using simple, interpretable models. The LIME methodology involves several key steps. First, for a given instance requiring explanation, LIME generates a set of perturbed samples by modifying the original feature values of that instance. These perturbed samples are then evaluated using the black-box model to obtain corresponding predictions. To maintain local fidelity, LIME assigns greater weights to perturbed samples that are closer to the original instance in feature space. A simple, interpretable model is subsequently trained on this weighted dataset, and the resulting learned coefficients of the surrogate model provide an approximation of feature importance.

Despite its popularity, LIME has several notable limitations. It often fails to capture discontinuous effects or effects that are prominent only within very narrow regions of the input space \citep{plumb2018model}. Furthermore, LIME generates perturbed instances by randomly modifying feature values, without accounting for correlations among features, which can lead to unrealistic or implausible data points \citep{sangroya2020guided}.

To address these limitations, we introduce the Locally Interpretable Individualized Treatment Rule (LI-ITR), a novel approach that constructs truly individualized treatment rules, in contrast to traditional methods that impose a universal rule. This design allows treatment decisions to vary across patients, thereby capturing nuanced treatment-response mechanisms. Additionally, LI-ITR is explicitly designed for interpretability, enhancing its clinical utility and practical applicability. Specifically, our approach leverages flexible models, such as neural networks, to capture complex treatment-response mechanisms, while generating locally interpretable models through patient-specific perturbed synthetic samples. To ensure these perturbations are realistic and preserve the underlying feature correlations, we adapt Variational Autoencoders (VAEs) \citep{kingma2013auto} to generate synthetic samples, thereby addressing one of the key limitations of existing methods such as LIME. Furthermore, we incorporate a mixture-of-linear-experts framework with a gating network \citep{jordan1994hierarchical} to construct patient-specific ITRs, balancing flexibility with transparency in individualized treatment decision-making. The gating network plays a central role by learning to assign patients to local expert models based on their individual characteristics, effectively selecting the most relevant local model for each subject. In contrast to LIME, which fits a single weighted model within a local neighborhood, our method improves robustness to the synthetic sample distribution by allowing dynamic selection among multiple local experts. This design not only mitigates the influence of poorly generated neighbors but also enables adaptation to heterogeneous local structures in the treatment-response landscape. Consequently, LI-ITR offers accurate and interpretable recommendations tailored to each patient's unique profile.

The remainder of the paper is structured as follows. In Section~\ref{Methodology}, we provide an in-depth explanation of the problem and our proposed LI-ITR method. Section~\ref{sim-stu} presents simulation studies to evaluate the performance of the LI-ITR method. In Section~\ref{data-ana}, we apply the LI-ITR method to identify optimal adjuvant endocrine therapy associated with a lower risk of hepatotoxicity. In Section~\ref{Disc}, we conclude with a discussion of the proposed method.

\section{Methodology}\label{Methodology}
Our proposed LI-ITR method operates in two stages. First, we generate realistic, instance-specific perturbations that preserve local data structure. Second, we model variability in these perturbations using a hierarchical mixture-of-experts \citep{jordan1994hierarchical}, yielding robust, subject-specific ITRs that capture nuanced local relationships while remaining interpretable.

\subsection{Problem Formulation}\label{prob_form}
We begin by introducing notation and defining the problem objective. Boldface denotes vectors, capital and lowercase letters denote random variables and their realizations, respectively.

Let \( Y \) denote the patient's outcome, which is assumed to be continuous, i.e., \( Y \in \mathbb{R} \) (e.g., biomarker levels), where higher values indicate better outcomes. Let \( \mathbf{X} = \{X_1, \ldots, X_p\}^{\mathrm{T}} \in \mathcal{X} \) denote the patient's observed feature vector, which may include genetic profiles, molecular biomarkers, phenotypic traits, psychosocial characteristics, and other relevant covariates. Additionally, let \( T \) represent the treatment variable, which can be either discrete (e.g., binary treatment assignment) or continuous (e.g., dosage levels of a drug). We define the full input vector as \( \mathbf{D} = \{T, \mathbf{X}^{\mathrm{T}}\}^{\mathrm{T}} \), capturing both treatment assignment and patient-specific covariates. Furthermore, let \( \mathbf{Z} = \{Z_1, \ldots, Z_{p^{'}}\}^{\mathrm{T}} \in \mathcal{Z} \) be a vector of latent variables representing a lower-dimensional representation of $\mathbf{X}$, capturing the underlying patterns and structure. Here, \( \mathcal{X} \) and \( \mathcal{Z} \) represent respective feature spaces of $\mathbf{X}$ and $\mathbf{Z}$, which may be high-dimensional in practice.

Assume the outcome is governed by the structural equation
\begin{equation*}
    Y = f(\mathbf{D}; \boldsymbol{\theta}) + \epsilon,
\end{equation*}
where \( f(\mathbf{D}; \boldsymbol{\theta}) \) is a complex, non-linear model mapping $\mathbf{D}$ to the outcome $Y$, $\boldsymbol{\theta}$ is the vector of learnable parameters, and \( \epsilon \) is an error term capturing unmeasured variability. Then, in a small neighborhood around a chosen data point, denoted \( \mathbf{d}_i \) ($i=1,\ldots, n$), \( f(\mathbf{D}; \boldsymbol{\theta}) \) can be approximated using a simple, interpretable model \( g_{d_i}(\mathbf{D}; \boldsymbol{\beta}_{d_i}) \). Specifically, for all \( \mathbf{D} \) in a sufficiently small neighborhood \( \mathcal{N}(\mathbf{d}_i) \), assume
\begin{equation}\label{local_eq}
f(\mathbf{D}; \boldsymbol{\theta}) \approx g_{d_i}(\mathbf{D}; \boldsymbol{\beta}_{d_i}), \quad \forall \mathbf{D} \in \mathcal{N}(\mathbf{d}_i)
\end{equation}
where \( g_{d_i}(\mathbf{D}; \boldsymbol{\beta}_{d_i}) \) can be a linear or low-degree polynomial model, chosen to provide a good local approximation to \( f(\mathbf{D}; \boldsymbol{\theta}) \). The choice of \( g_{d_i}(\mathbf{D}; \boldsymbol{\beta}_{d_i}) \) depends on the complexity of \( f(\mathbf{D}; \boldsymbol{\theta}) \) and the desired level of interpretability.

Estimating the local model \( g_{d_i}(\mathbf{D}; \boldsymbol{\beta}_{d_i}) \) can be approached in two distinct ways. The first approach relies directly on the observed data and involves identifying the appropriate neighborhood or implementing a weighted regression to construct the interpretable ITR. However, the effectiveness of this approach depends critically on the choice and size of the neighborhood, as well as on the complexity of the underlying local structure in the treatment-response relationship. This issue is amplified in high-dimensional settings, where data sparsity due to the curse of dimensionality becomes more severe. Moreover, limited or restricted access to training data due to data-sharing limitations further hampers the reliable estimation of local models.

The other approach to estimating the local model \( g_{d_i}(\mathbf{D}; \boldsymbol{\beta}_{d_i}) \) reformulates the problem as approximating the behavior of a trained black-box model locally, rather than directly approximating \( f(\mathbf{D}; \boldsymbol{\theta}) \) using observed data. That is, we rewrite (\ref{local_eq}) as
\begin{equation*}
\hat{Y} \approx g_{d_i}(\mathbf{D}^{'}; \boldsymbol{\beta}_{d_i}), \quad \forall \mathbf{D^{'}} \in \mathcal{N}(\mathbf{d}_i),
\end{equation*}
where $\hat{Y}=f(\mathbf{D}^{'}; \hat{\boldsymbol{\theta}})$ and \( \mathbf{D}^{'} \) is generated by perturbing \( d_i \). This approach alleviates data sparsity and eliminates the need for direct access to the training data. In this case, the reliability of the estimated local model depends on the black-box model's accuracy in estimating $f(\mathbf{D}; \boldsymbol{\theta})$, the realism of perturbed samples, and the effectiveness of the local model estimation method. While black-box accuracy is relevant, our focus is on developing a principled methodology for generating realistic or near-realistic synthetic perturbations and estimating \( g_{d_i}(\mathbf{D}^{'}; \boldsymbol{\beta}_{d_i}) \) to produce stable, interpretable, and subject-specific ITRs.

\subsection{Model Framework}\label{Model_frame}
Our methodology consists of four steps: (1) train a flexible black-box model, (2) generate perturbed synthetic samples for the given subject of interest, (3) obtain predictions for these samples from the trained black-box model, and (4) use the perturbed samples and predictions to estimate a local model for interpretable ITRs. Below, we elaborate on steps 2 and 4, which form the core of our approach.

\subsubsection{Generating Perturbed Synthetic Samples}\label{gen_pert_sam}
Generating perturbed synthetic samples is essential for constructing reliable local models, as it reveals how a black-box model responds to small variations in the input and clarifies its local behavior. Synthetic data have also been shown to address key challenges in ML, such as limited sample sizes and privacy concerns. Recent studies demonstrate that high-quality synthetic data can improve model performance, preserve privacy, and enhance diagnostic accuracy \citep{shen2023boosting, ghalebikesabi2023differentially, gao2023synthetic}.

Realism is critical when generating synthetic samples, as only plausible variations yield reliable local models. Independent feature perturbation, as used in LIME, is computationally convenient but disrupts natural feature correlations, often producing unrealistic samples. To address this limitation, we propose a tailored VAE architecture that captures and preserves the joint distribution of input features, generating realistic samples that respect these underlying correlations. The VAE simultaneously preserves feature dependencies and performs dimensionality reduction via a probabilistic encoder-decoder structure.

Assume that $\mathbf{X}$ follows the distribution $p\left(\mathbf{X}\right)$, and that this distribution is described by some underlying latent variables $\mathbf{Z}$ which follow some prior distribution $p\left(\mathbf{Z}\right)$. The joint distribution $p(\mathbf{X}, \mathbf{Z})$ can be expressed as follows:
\begin{align*}
    p(\mathbf{X}, \mathbf{Z})=p_{\boldsymbol{\gamma}}(\mathbf{X} \mid \mathbf{Z}) p(\mathbf{Z}),
\end{align*}
with $p_{\boldsymbol{\gamma}}(\mathbf{X} \mid \mathbf{Z})$ representing the probabilistic decoder. The prior distribution \( p(\mathbf{Z}) \) is typically chosen to be a multivariate zero-centered normal distribution with unit covariance, i.e., \( \mathbf{Z} \sim \mathcal{N}(\mathbf{0}, \mathbf{I}) \) \citep{kingma2013auto}. The posterior distribution $p_{\boldsymbol{\gamma}}\left(\mathbf{Z} \mid \mathbf{X}\right)$ is analytically intractable and is therefore approximated using a tractable distribution $q_{\boldsymbol{\phi}}\left(\mathbf{Z} \mid \mathbf{X}\right)$. This approximating distribution, known as the probabilistic encoder, encodes the input data $\mathbf{X}$ into the latent space. To encourage more disentangled and structured latent representations, a modification known as $\beta$-VAE \citep{higgins2017beta} introduces a scaling factor $\beta$, allowing explicit control over the trade-off between reconstruction fidelity and latent space regularization. The objective of the $\beta$-VAE algorithm is to maximize the modified evidence lower bound (modified-ELBO), which is expressed as
\begin{equation}\label{VAE-obj}
    \text{modified-ELBO} = \mathbb{E}_{q_{\boldsymbol{\phi}}\left(\mathbf{Z} \mid \mathbf{X}\right)}\left[\log p_{\boldsymbol{\gamma}}(\mathbf{X} \mid \mathbf{Z})\right] - \beta \cdot \text{KL}\left(q_{\boldsymbol{\phi}}\left(\mathbf{Z} \mid \mathbf{X}\right) \parallel p(\mathbf{Z})\right).
\end{equation}
In (\ref{VAE-obj}), the first term is the reconstruction error for $\mathbf{X}$ given $\mathbf{Z}$, and \(\text{KL}\left(q_{\boldsymbol{\phi}}\left(\mathbf{Z} \mid \mathbf{X}\right) \parallel p(\mathbf{Z})\right)\) represents the Kullback–Leibler (KL) divergence \citep{cover1999elements}, measuring the difference between the approximating posterior distribution $q_{\boldsymbol{\phi}}\left(\mathbf{Z} \mid \mathbf{X}\right)$ and the prior distribution $p(\mathbf{Z})$.

\begin{figure}[ht!]
\centering
\includegraphics[width=7cm,height=16cm]{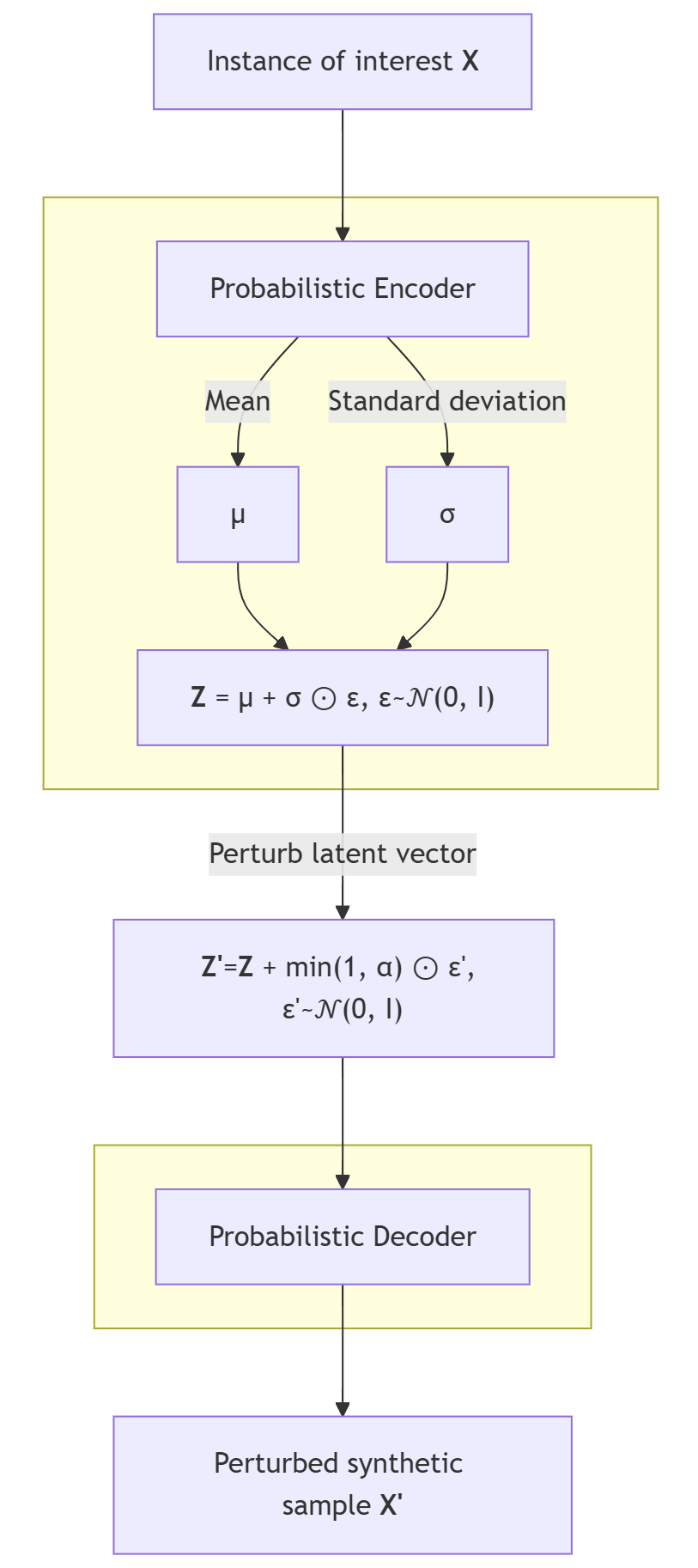}
\caption{Illustration of the proposed strategy for generating perturbed synthetic samples}
\label{vae-perturb-flowchart}
\end{figure}

This objective function encourages an independent latent structure by aligning the posterior distribution of $\mathbf{Z}$ with the standard normal prior, while simultaneously ensuring accurate reconstruction of the original data. Instead of perturbing samples in the original, potentially correlated feature space, we perturb the decorrelated latent variables $\mathbf{Z}$. To be specific, we generate perturbed latent variables, denoted $\mathbf{Z}^{'}$, as follows:
$$\mathbf{Z}^{'}=\mathbf{Z} + \min(1, \alpha) \odot \boldsymbol{\epsilon}^{'}, \quad \boldsymbol{\epsilon}^{'} \sim \mathcal{N}(\mathbf{0}, \mathbf{I}),$$
where $\alpha \in (0, 1)$ can either be set as a fixed scalar or allowed to increase. This enables free perturbation of an instance's latent variables and generation of synthetic samples via the probabilistic decoder, producing instance-relevant samples while reducing the risk of unrealistic ones (Figure \ref{vae-perturb-flowchart}). Unlike the feature vector $\mathbf{X}$, the treatment assignment $T$ is not an input to the VAE. Instead, we invoke the \textit{positivity} condition commonly adopted in causal inference \citep{hernan2010causal} to ensure that both treatment regimes are well-defined across the covariate space. We therefore generate a synthetic binary treatment variable $T^{'}$ by sampling from a Bernoulli distribution with probability $0.5$. We concatenate $T^{'}$ with the perturbed synthetic feature vector, denoted $\mathbf{X}^{'}$, to form the complete perturbed synthetic sample, denoted $\mathbf{D}^{'}$, and let $m$ denote the total number of perturbed synthetic samples.

\subsubsection{Estimating Subject-Specific Treatment Rules}\label{learn-alg}
After generating instance-specific $\mathbf{D}^{'}$, we estimate a local model using the perturbed synthetic samples. However, a single local model may be insufficient, as the perturbed neighborhood can be dispersed and may reflect heterogeneous or even opposing treatment–response patterns rather than a single coherent local structure. This limitation is particularly relevant in high-dimensional settings, where local linear approximations—such as those employed by LIME—implicitly assume that the black-box model can be well represented by a single surrogate within a predefined neighborhood. To address this limitation, we frame the local model estimation as a clustering problem where we assume \( \hat{Y} \) is generated from a mixture of $K$ components, each corresponding to a distinct local regime. Specifically, we model \( \hat{Y} \) as arising from a mixture of distributions \(\mathcal{D}(\mu_{k}(\mathbf{D}^{'}, \boldsymbol{\beta}_{k}), \sigma_k)\), where the mean function $\mu_{k}(\mathbf{D}^{'}, \boldsymbol{\beta}_{k})$ is specified as a linear function of $\mathbf{D}^{'}$ incorporating both main and interaction effects, and $\sigma_k$ denotes the associated scale parameter. Specifically, we specify
\begin{equation}\label{mean_model}
    \mu_{k}(\mathbf{D}^{'}, \boldsymbol{\beta}_{k}) = \boldsymbol{\beta}_{k1}^{\mathrm{T}} H_{0} + (\boldsymbol{\beta}_{k2}^{\mathrm{T}} H_{1}) T^{'},
\end{equation}
where $\boldsymbol{\beta}_{k1}$ and $\boldsymbol{\beta}_{k2}$ are the associated regression coefficients, and we consider two covariate subsets: \( H_{0} \), which includes prognostic variables that directly influence the outcome regardless of treatment, and \( H_{1} \), which contains \textit{prescriptive} or \textit{tailoring} variables that modify the treatment effects \citep{chakraborty2013statistical}. Both \( H_{0} \) and \( H_{1} \) may include an intercept term and can share overlapping covariates. The objective is to estimate the parameters in (\ref{mean_model}), and estimate the optimal treatment determined by:
\begin{align*}
    d_{k}=\arg\underset{t}{\max} \ \boldsymbol{\beta}_{k1}^{\mathrm{T}} H_{0} + (\boldsymbol{\beta}_{k2}^{\mathrm{T}} H_{1}) t,
\end{align*}
with $t$ denoting the realization of the random variable $T$. It is noted that the prognostic part in (\ref{mean_model}), $\boldsymbol{\beta}_{k1}^{\mathrm{T}} H_{0}$, does not need to be interpretable and can be left unspecified, allowing for its approximation using flexible black-box models.

To this end, for the $j$-th perturbed synthetic sample, we have
\[
p(\hat{Y}_j \mid \mathbf{D}_j^{'}) = \sum_{k=1}^K \pi_{jk}(\mathbf{X}_j^{'}; \mathbf{w}) \, \mathcal{D}\left(\hat{Y}_j\mid \mu_{k}(\mathbf{D}_j^{'}, \boldsymbol{\beta}_{k}), \sigma_k\right),\qquad j=1, \dots, m,
\]
where $\pi_{jk}(\mathbf{X}_j^{'}; \mathbf{w})$ represents the responsibility of the $k$-th distribution in modeling the $j$-th perturbed synthetic sample. To be specific, we let
\begin{equation}\label{respons_eq}
    \pi_{jk}(\mathbf{X}_j^{'}; \mathbf{w}) = \frac{\exp\left( h_k(\mathbf{X}_j^{'}; \mathbf{w}_k) \right)}{\sum\limits_{l=1}^{K} \exp\left( h_l(\mathbf{X}_j^{'}; \mathbf{w}_l) \right)},\quad \text{for}\quad k=1,\dots,K,
\end{equation}
where \( h_k : \mathbb{R}^p \to \mathbb{R} \) is a differentiable function with parameter $\mathbf{w}_k \subset \mathbf{w}$. We let $\pi_{j}(\mathbf{X}_j^{'}; \mathbf{w})\coloneqq \big\{\pi_{j1}(\mathbf{X}_j^{'}; \mathbf{w}),\dots,\pi_{jK}(\mathbf{X}_j^{'}; \mathbf{w})\big\}^{\mathrm{T}}$ denote the assignment probability vector for subject $j$. The objective is to find parameters \(\boldsymbol{\eta} \coloneqq \{\mathbf{w}_k, \boldsymbol{\beta}_{k}, \sigma_k\}_{k=1}^K \) that maximizes the log-likelihood
\begin{equation}\label{main_log_lik}
    \mathcal{L}(\boldsymbol{\eta}) = \sum_{j=1}^m \log\ p(\hat{Y}_j \mid \mathbf{D}_j^{'}).
\end{equation}

We use a single layer neural network, referred to as the \textit{expert}, to represent the interpretable linear model (\ref{mean_model}), and a multilayer neural network, referred to as the \textit{gating network}, to model $h_k(\mathbf{X}_j^{'}; \mathbf{w}_k)$ in (\ref{respons_eq}). This allows non-linear, flexible mappings for distribution assignments. Model parameters are optimized jointly using standard iterative methods such as AdamW \citep{loshchilov2019decoupled}, which simultaneously update the expert-specific parameters as well as the parameters involved in (\ref{respons_eq}) for expert assignment.

We note that direct optimization of (\ref{main_log_lik}) can result in soft expert assignments, whereby each sample may be partially assigned to multiple experts depending on the gating network output. While such soft assignments are common in mixture models, they can hinder the interpretability of the resulting local models and introduce bias in parameter estimation, particularly when the gating network is prone to identifying shortcut solutions that dominate expert selection. To address these challenges, we propose two unique model structures for learning the local experts. Firstly, we introduce a negative Shannon entropy \citep{shannon1948mathematical} penalty to encourage sparse expert selection. This modifies the objective function (\ref{main_log_lik}) as follows:
\begin{equation*}
    \mathcal{L}(\boldsymbol{\eta}) = \sum_{j=1}^m \log\ p(\hat{Y}_j \mid \mathbf{D}_j^{'}) + \lambda \sum_{j=1}^{m} \pi_{j}(\mathbf{X}_j^{'}; \mathbf{w})^\mathrm{T} \log \pi_{j}(\mathbf{X}_j^{'}; \mathbf{w}),
\end{equation*}
where $\lambda>0$ controls the strength of the sparsity-inducing penalty. By discouraging diffuse expert assignments, this regularization improves the interpretability of local surrogate models and stabilizes parameter estimation by mitigating degenerate gating behavior. While larger values of $\lambda$ encourage sparse expert distribution, the resulting assignments may still remain soft. To obtain fully interpretable local models, we therefore impose an additional hard selection step on the gating network output, forcing each synthetic sample to be assigned to a single expert. Specifically, we define $\tilde{\pi}_{j}(\mathbf{X}_j^{'}; \mathbf{w}) = \mathbf{e}_{\arg\max_k \pi_{jk}(\mathbf{X}_j^{'}; \mathbf{w})}$ to be the output of the gating network, where \( \mathbf{e}_i \in \mathbb{R}^K \) denotes the standard basis vector with a $1$ at the \( i \)-th position and $0$ elsewhere. Since the \( \arg\max \) function is non-differentiable, we use the Straight-Through Estimator (STE) to enable gradient-based optimization as follows:
\begin{align*}
    \tilde{\pi}_{j}(\mathbf{X}_j^{'}; \mathbf{w}) = \texttt{StopGradient}(\mathbf{e}_{\arg\max_k \pi_{jk}(\mathbf{X}_j^{'}; \mathbf{w})} - \pi_{j}(\mathbf{X}_j^{'}; \mathbf{w}))+\pi_{j}(\mathbf{X}_j^{'}; \mathbf{w}),
\end{align*}
where \texttt{StopGradient} denotes the use of a stop-gradient (or \texttt{detach}) operation, which ensures that the hard selection is used in the forward pass, while gradients are propagated through the softmax distribution during backpropagation.

Once the model is trained, we determine the respective expert for the subject of interest by feeding its baseline input to the trained gating network, and the estimated parameters of the selected expert are used to develop a treatment rule. To be specific, let $\hat{\boldsymbol{\beta}}_{k1}^{\mathrm{T}}$ and $\hat{\boldsymbol{\beta}}_{k2}^{\mathrm{T}}$ denote the estimated regression coefficients of the selected expert, and the optimal treatment, denoted $\hat{d}_k$, is determined by:
\begin{align*}
    \hat{d}_{k}=\arg\underset{t}{\max} \ \hat{\boldsymbol{\beta}}_{k1}^{\mathrm{T}} H_{0} + (\hat{\boldsymbol{\beta}}_{k2}^{\mathrm{T}} H_{1}) t.
\end{align*}

\section{Simulation Studies}\label{sim-stu}
\subsection{Simulation Designs}\label{sim-design}
We simulate a globally nonlinear outcome function that can be locally approximated by simple linear models within predefined regions of the input space, reflecting practical settings with complex global structure and interpretable local behavior. Specifically, we first generated four baseline covariates $\{X_1, X_2, X_3, X_4\}$ from a 1-dimensional latent variable, $Z \sim U(0, 20)$. The mean structures were defined as $\mu_{X_1} = 1.50 \sin(Z)$, $\mu_{X_2} = 1.25 \cos(Z)$, $\mu_{X_3} = 1.65 \sin(Z)$, and $\mu_{X_4} = 1.25 \cos(Z)$. The covariates were then generated as follows; $X_1 \sim \text{LogNormal}(\mu_{X_1}, 0.05)$, $X_2 \sim \mathcal{N}(\mu_{X_2}, 0.55)$, $X_3 \sim \mathcal{N}(\mu_{X_3}, 0.65)$, $X_4 \sim \text{LogNormal}(\mu_{X_4}, 0.05)$.

Treatment assignment probability was modeled using a logistic function, \( p(T = 1 \mid X_3, X_4) = \exp(-0.65 X_3 + 0.15 X_4)/ \big(1 + \exp(-0.65 X_3 + 0.15 X_4)\big) \), and treatment \(T\) was drawn from a Bernoulli distribution, \( T \sim \text{Bernoulli}\big(p(T = 1 \mid X_3, X_4)\big) \).

The outcome variable $Y$ was generated using a piecewise heterogeneous treatment effect model. Let $x_{1}^{\text{med}}$ and $x_{2}^{\text{med}}$ denote the empirical medians of $X_1$ and $X_2$, respectively. Then for each observation $i = 1, \dots, n$:
\begin{equation*}
    Y_i = f(X_{1i}, X_{2i}, X_{3i}, X_{4i}, T_i) + \varepsilon_i,
\end{equation*}
where $\varepsilon_i$ is the error term, and we defined four submodels indexed by \( k \in \{1, 2, 3, 4\} \), corresponding to regions of the input space partitioned by median splits on \( X_1 \) and \( X_2 \). To be specific, $f(\cdot)$ was defined as:
\[
f = \boldsymbol{\beta}_{1}^{\mathrm{T}} H_{0} + \left(\boldsymbol{\beta}_{k2}^{\mathrm{T}} H_{1}\right) T,
\]
where $H_0 = \{X_1, X_2, X_3, X_4\}^{\mathrm{T}}$, and $H_1 = \{1, X_1, X_2\}^{\mathrm{T}}$. The main effect coefficients were constant across all subregions; i.e., $\boldsymbol{\beta}_{1} \coloneqq (\beta_{11}, \beta_{21}, \beta_{31}, \beta_{41})^{\mathrm{T}} = (2.25, 1.65, 1.55, 1.25)^{\mathrm{T}}$. However, $\boldsymbol{\beta}_{k2} \coloneqq (\beta_{1k2}, \beta_{2k2}, \beta_{3k2})^{\mathrm{T}}$ varied by median splits of \( X_1 \) and \( X_2 \), with the corresponding coefficient vectors reported in the Supplementary Material.

Training data were generated under three sample size settings, \( n = 40,000 \), \( 10,000 \), and \( 2,000 \), and a test set of $1,000$ instances was used across all scenarios. For each setting, perturbed synthetic samples of sizes \(m = 800,000\) and \(100,000\) were generated.

To assess the robustness of LI-ITR to misspecification in the prescriptive component (i.e., $\boldsymbol{\beta}_{k2}^{\mathrm{T}} H_{1}$) of the model (\ref{mean_model}), we modified the data-generating process by introducing a quadratic term (full details provided in the Supplementary Material).

\subsection{Simulation Results}\label{sim-results}
Table~\ref{tab12M} presents the average biases of $\boldsymbol{\beta}_{1}$ and $\boldsymbol{\beta}_{k2}$, along with the corresponding standard deviations. Across all training sample sizes, both LI-ITR and LIME exhibit low bias and modest variability for coefficients associated with baseline covariates (\(X_1-X_4\)), indicating robust recovery of common covariate effects that do not vary across subjects. For LI-ITR, biases are close to zero, and variability generally decreases with increasing training and synthetic sample sizes.

In contrast, substantial differences emerge for treatment-related coefficients \(\beta_{1k2}\), \(\beta_{2k2}\), and \(\beta_{3k2}\), which govern subject-specific treatment effects. LI-ITR consistently yields low bias (below $0.10$ across all settings) and decreasing variability with the increase of the training and synthetic sample sizes. LIME, however, exhibits substantially larger bias and variability for these parameters, with little improvement from larger training or synthetic sample sizes.

These findings demonstrate a clear difference in performance between the two methods, particularly for parameters involved in treatment rule specification. LI-ITR provides reliable estimates for the coefficients governing the individualized treatment effect, while LIME produces unacceptably biased and highly variable estimates.

Furthermore, to evaluate the performance of LI-ITR in identifying optimal treatments on the test data, we compared it with LIME and three established approaches: OWL \citep{zhao2012estimating}, causal forest \citep{wager2018estimation}, and Q-learning \citep{chakraborty2013statistical}, as well as the treatment rule induced directly by the black-box model, referred to as the \textit{black-box ITR}. Accuracy is quantified by the proportion of recommended treatments matching the true optimal treatments. As shown in Table \ref{tab:pcot_results_cor_quad}, LI-ITR consistently achieves accuracy rates above $99\%$ across all scenarios, showing nearly identical performance to that of black-box ITR and outperforming all other methods. Causal forest also performs well with accuracy rates ranging from $97\%$ to $99\%$. In contrast, the LIME method shows substantially lower accuracy at approximately $88\%$, while OWL and Q-learning perform poorly with accuracy rates around $83\%$. Notably, the LI-ITR method's performance appears robust to sample size variations, suggesting efficient learning even with smaller training and synthetic sample sizes.

Results under the misspecified expert models, also reported in Table \ref{tab:pcot_results_cor_quad}, indicate that LI-ITR, black-box ITR, and causal forest achieve the best overall performance, with LI-ITR remaining highly robust to model misspecification ($>0.98$) across all settings despite model misspecification. Moreover, LI-ITR uniquely provides subject-level interpretability for treatment decisions, a desirable feature that causal forests do not offer. In contrast, Q-learning and LIME exhibit stable but inferior performance, while OWL shows increasing instability as sample size decreases.

\section{Real Data Application}\label{data-ana}
Aspartate aminotransferase (AST) is a widely used biomarker for monitoring hepatocellular injury in patients receiving endocrine therapy for breast cancer. Serum elevation of this enzyme generally reflects hepatocellular damage \citep{giannini2005liver}. Accordingly, clinical guidelines recommend routine AST monitoring, as early detection of hepatotoxicity enables timely intervention and prevents progression to serious liver injury.

While adjuvant endocrine therapies have well-established efficacy in reducing breast cancer recurrence \citetext{\citetalias{breast2005comparison}}, their hepatotoxic effects remain a clinical concern that varies substantially across patients. For many patients, multiple endocrine therapies offer similar long-term survival, yet may differ meaningfully in their propensity to induce liver enzyme elevation \citetext{\citetalias{early2015aromatase}; \citealp{lin2014prospective}}. In such settings, treatment selection should be guided not only by efficacy, but also by the patient-specific risk of hepatotoxicity. Our goal is therefore to achieve precision control of treatment-induced toxicity by identifying for each patient, the endocrine therapy that minimizes expected AST elevation conditional on baseline liver function and other baseline characteristics, as well as laboratory measurements. Because the relationship between treatment, patient characteristics, and AST response is often highly nonlinear and heterogeneous, accurately capturing this risk requires flexible ML models. At the same time, treatment decisions driven by hepatotoxicity must remain transparent and clinically interpretable, since they directly affect monitoring intensity, dose modification, and treatment switching. This setting naturally motivates the use of locally interpretable individualized treatment rules, which combine the predictive power of modern ML with patient-specific, human-readable treatment recommendations.

To this end, we focus on $1,819$ women with hormone–receptor–positive breast cancer who initiated adjuvant endocrine therapy between January 13, 2016, and November 29, 2023. Among them, $n=445$ ($24.5\%$) received tamoxifen, and $n=1,374$ ($75.5\%$) received an aromatase inhibitor (AI) therapy. Patient baseline characteristics are summarized in the Supplementary Material.

Pre-treatment laboratory measurements included alkaline phosphatase (ALP), alanine aminotransferase (ALT), AST, absolute neutrophil count, white blood cell (WBC) count, blood urea nitrogen (BUN), serum creatinine, total bilirubin, and estimated glomerular filtration rate (eGFR). Pre-treatment values were obtained within eight weeks prior to treatment initiation, with the measurement closest to the start of therapy selected for each patient. Summary statistics for the pre-treatment laboratories are provided in Table~\ref{baseline_lab}. Post-treatment AST measurements were collected between four and twelve weeks following treatment initialization ($28-84$ days post-treatment), again selecting the measurement closest to the end of the treatment window. The primary toxicity endpoint was defined as the change in AST, calculated as the post-treatment value minus the pre-treatment value. The mean post-treatment AST was $23.6 \pm 14.4$ U/L (median: $20$; IQR: $17-26$; range: $5-303$). Among $1,819$ patients, the change in AST from baseline to post-treatment had a mean decrease of $1.0$ U/L (SD $22.6$), with values ranging from a decrease of $731$ U/L to an increase of $258$ U/L, reflecting substantial inter-individual variability in treatment-induced hepatotoxicity.

To implement the LI-ITR method, we first trained a neural network to predict changes in AST. The model input included pre-treatment laboratory measurements, patient baseline characteristics, and treatment information. The training/test split was performed in a stratified manner based on the treatment variable, resulting in $1,364$ samples for training and $455$ samples for testing, while preserving the distribution of treatment classes. A detailed description of the neural network architecture can be found in the Supplementary Material.

The neural network model yields a coefficient of determination ($R^2$) of $0.91$ on the test data, substantially outperforming the linear model ($R^2=0.78$), underscoring the need for a more flexible modeling approach to characterize treatment-response relationships.

We then trained a VAE model to learn a lower-dimensional latent representation of the pre-treatment laboratory and baseline features, where the model architecture and model training are detailed in the Supplementary Material.

The correlation structure in the test data reveals markedly weaker dependence patterns in the derived latent variables compared to the original features, as shown in Figures~S1 and S2 of the Supplementary Material. This reduced dependence enables more realistic and coherent local perturbations in latent space than perturbations applied directly to the highly correlated original features, as in LIME.

Next, we estimated interpretable local models for each subject in the test set, except for one subject with very extreme values that caused the perturbed samples to fall outside the training data range. To mitigate the impact of multicollinearity on both interpretability and numerical stability, we imposed $\ell_{2}$-regularization (weight decay) on the local models and tuned the regularization strength through repeated cross-validation as detailed in the Supplementary Material.

Using the estimated local models, we identified the treatment that yields the least elevation in AST, and subsequently calculated the value function,
\begin{equation*}
    V(\hat{d}_{k}) = \frac{1}{n} \sum_{i=1}^{n} \frac{\mathbb{I}(T_i = \hat{d}_{k}) \cdot Y_i}{\hat{e}(\mathbf{X}_i) T_i + (1 - \hat{e}(\mathbf{X}_i))(1 - T_i)},
\end{equation*}
where $\mathbb{I}(\cdot)$ is the indicator function for a treatment match, and $\hat{e}(\mathbf{X}_i)$ is the propensity score estimated using a Probability Forest \citep{athey2019generalized}, a non-parametric ensemble method that extends the random forest framework to estimate conditional class probabilities $P(T_i = 1 \mid \mathbf{X}_i)$.

Table~\ref{tab:treatment_dist_policy} summarizes the treatment allocation patterns and the corresponding estimated policy values for all methods considered. The proposed LI-ITR recommends an AI for $28$ of the $454$ patients ($6\%$), assigning the remaining $94\%$ to tamoxifen. This prescription rate is markedly lower than that of Q-learning, which assigns AI to $19\%$ of patients, and only slightly lower than the rates produced by black-box ITR ($7\%$) and LIME ($7\%$). The causal-forest model is the most conservative, prescribing tamoxifen universally, whereas the OWL approach recommends AI for $9\%$ of patients.

Policy performance is assessed through the estimated value function, where more negative values indicate lower expected toxicity and greater clinical benefit under the proposed policy. LI-ITR achieves the most favorable value ($-2.922$), outperforming Q-learning ($-2.196$), OWL ($-2.403$), causal forest ($-2.733$), LIME ($-2.413$), and black-box ITR ($-2.800$). Notably, LI-ITR achieves these gains while remaining highly parsimonious and interpretable. Compared with LIME, LI-ITR provides a substantially more faithful approximation to the underlying neural network model. Across $454$ subjects, the mean absolute difference between LI-ITR's prediction and the black-box prediction was $0.17$ (SD $=0.24$), whereas the corresponding error for LIME was more than three times larger at $0.54$ (SD $=0.61$). These results highlight the benefit of advancing beyond LIME by coupling a VAE-based perturbation strategy with gating-network–driven local model learning.

\begin{figure}[ht!]
\includegraphics[width=16.5cm,height=9cm]{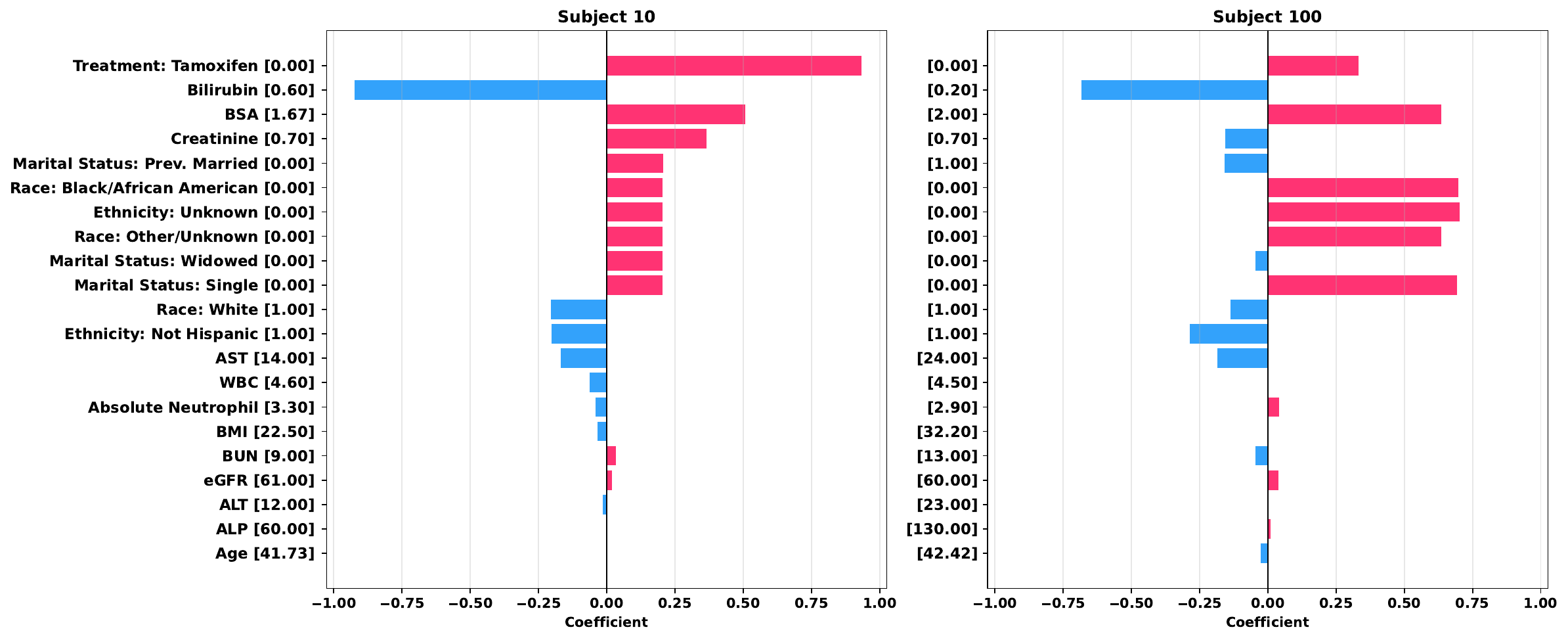}
\caption{Estimated treatment and treatment–feature interaction effects for two subjects. Values in brackets $[\cdot]$ denote the corresponding feature values}
\label{10to100_comparison}
\end{figure}

We then examined heterogeneity in fitted coefficients across subjects. As shown in Figures~S3 and S4 of the Supplementary Material, both continuous and categorical features reveal pronounced inter-individual variation in estimated effects. To further illustrate this contrast, we examined two test-set subjects (Subjects 10 and 100). For both, the local models achieve high fidelity to the black-box predictor, with coefficients yielding predictions that closely match the corresponding black-box outputs (local $R^2 = 0.96$ and $0.97$, respectively). This close agreement confirms that the extracted coefficients provide faithful local approximations of the black-box behavior. Comparison of their individualized coefficient vectors (Figure~\ref{10to100_comparison}), capturing treatment and treatment-feature interactions, reveals markedly different local structures despite some overlap in active features. For Subject~10, the tamoxifen main effect is large and positive, with laboratory interactions dominating the local approximation: total bilirubin has a strong negative coefficient, while creatinine shows a substantial positive effect. Even though this subject's laboratory values fall within clinically normal ranges, these coefficients indicate heightened local sensitivity to hepatic and renal markers.

In contrast, Subject~100 shows a much smaller main effect for tamoxifen and a different pattern of interaction effects. Demographic variables, including race and marital status, exhibit larger coefficients, while multiple laboratory interactions differ in both sign and scale relative to Subject~10 (e.g., creatinine and BUN shift from positive to negative effects). These differences are expected given Subject~100's higher BMI, elevated liver enzyme values, and distinct marital status, indicating that the black-box model relies on a different combination of predictors in this local neighborhood.

Overall, these results highlight the limitations of a single interpretable treatment rule and underscore the value of subject-specific treatment policies. By fitting individualized local models, LI-ITR yields interpretable, high-fidelity approximations that reveal how the black-box rule adapts across subjects with heterogeneous clinical and demographic characteristics.

\section{Discussion}\label{Disc}
In this study, we introduce the LI-ITR method to facilitate the development of interpretable ITRs from any black-box model, with the goal of recovering high-fidelity, subject-specific decision rules that remain transparent and clinically interpretable. By moving beyond traditional one-size-fits-all treatment policies, LI-ITR enables individualized treatment rules that adapt to patient-specific clinical and demographic characteristics. Simulation studies demonstrate that the LI-ITR method faithfully approximates the black-box behavior while accurately recovering true local coefficients and optimal treatment decisions. Moreover, application to real-world data highlights the necessity and utility of LI-ITR for identifying subject-specific treatment policies while preserving interpretability.

Although we adopt a standard VAE algorithm to generate realistic perturbed samples, alternative formulations may be advantageous in settings with strongly correlated features, in particular, a Total Correlation Variational Autoencoder (TCVAE) \citep{chen2018isolating}. TCVAE decomposes the standard VAE KL divergence term into three interpretable components: (i) a mutual information term, (ii) a total correlation term, and (iii) a dimension-wise KL. By placing a greater weight on the total correlation component, TCVAE encourages the disentanglement of correlated structures in the data, yielding latent representations that are closer to statistically independent.

An important practical consideration in LI-ITR is the expressive power of the local model used to approximate the black-box predictor. Simulation results suggest that LI-ITR is relatively robust to moderate misspecification of this model; nevertheless, its validity rests on two conditions: the neighborhood around the target observation must be chosen small enough, and second, within this restricted region, the black-box response surface should be well approximated by the first-order term of its Taylor expansion. If higher-order structure is present, richer surrogate families can be adopted. In all cases, the adequacy of the chosen local model can be quantified with standard goodness-of-fit criteria such as $R^2$.

A second practical concern is \emph{multicollinearity} among predictors, a phenomenon that is not unique to our model. Multicollinearity can adversely affect both the interpretability and the numerical stability of LI-ITR's local models. To address this issue, we incorporated $\ell_2$-regularization (weight decay) into the local linear models and selected the regularization strength using repeated cross-validation. This strategy provided effective shrinkage of correlated coefficients while preserving high-fidelity local approximations.

More broadly, LI-ITR allows flexible modeling of the prognostic component of the local outcome model while maintaining interpretability of treatment effects. In our real-data application, we adopted a \emph{two-head} architecture that explicitly decomposes the baseline response and treatment-related effects. This design offers two practical advantages: it allows greater flexibility in modeling the baseline response while keeping the treatment-effect component parsimonious and interpretable, and it enables the use of differential regularization schemes across components, which can empirically improve numerical stability and reduce parameter redundancy in practice.

\section*{Acknowledgments}
\textit{This work was supported in part by the National Cancer Institute (Grant Nos. P30 CA008748, P50 CA271357, P50 CA271357-02S1, and R25 CA272282).}



\bibliographystyle{apalike}
\bibliography{sn-bibliography}

\newpage




\begin{table}[ht]
\centering
\resizebox{0.27\textheight}{!}{
\rotatebox{90}{%
\begin{threeparttable}
\caption{Bias and standard deviation for $\boldsymbol{\beta}_1$ and $\boldsymbol{\beta}_{k2}$ for two synthetic sample sizes}
\label{tab12M}
\begin{tabular}{llccccccccccccccccccccc}
\toprule
& & \multicolumn{7}{c}{$n=2{,}000$} & \multicolumn{7}{c}{$n=10{,}000$} & \multicolumn{7}{c}{$n=40{,}000$} \\
\cmidrule(lr){3-9}\cmidrule(lr){10-16}\cmidrule(lr){17-23}
$m$ & &
$\beta_{11}$ & $\beta_{21}$ & $\beta_{31}$ & $\beta_{41}$ & $\beta_{1k2}$ & $\beta_{2k2}$ & $\beta_{3k2}$ &
$\beta_{11}$ & $\beta_{21}$ & $\beta_{31}$ & $\beta_{41}$ & $\beta_{1k2}$ & $\beta_{2k2}$ & $\beta_{3k2}$ &
$\beta_{11}$ & $\beta_{21}$ & $\beta_{31}$ & $\beta_{41}$ & $\beta_{1k2}$ & $\beta_{2k2}$ & $\beta_{3k2}$ \\
\midrule
\multirow{6}{*}{$100{,}000$}
    & \multicolumn{21}{l}{\textbf{LI-ITR}} \\
    & Bias & 0.0758 & 0.0100 & 0.0060 & 0.0308 & 0.0403 & 0.0845 & 0.0893 & 0.0545 & 0.0068 & 0.0039 & 0.0531 & 0.0277 & 0.0440 & 0.0913 & 0.0395 & 0.0037 & 0.0017 & 0.0457 & 0.0119 & 0.0436 & 0.0606 \\
    & SD & 0.1893 & 0.0192 & 0.0111 & 0.0657 & 0.0677 & 0.2793 & 0.4112 & 0.1065 & 0.0112 & 0.0045 & 0.1165 & 0.0482 & 0.1800 & 0.4455 & 0.0900 & 0.0055 & 0.0015 & 0.1201 & 0.0289 & 0.2538 & 0.3745 \\[1ex]    
    & \multicolumn{21}{l}{\textbf{LIME}} \\
    & Bias & 0.0262 & 0.0048 & 0.0295 & 0.0090 & 0.7185 & 1.4696 & 0.5760 & 0.0229 & 0.0149 & 0.0131 & 0.0170 & 0.6308 & 1.4868 & 0.5353 & 0.0076 & 0.0067 & 0.0071 & 0.0083 & 0.7166 & 1.4844 & 0.5228 \\
    & SD & 0.0171 & 0.0049 & 0.0174 & 0.0087 & 0.3301 & 1.2394 & 0.3778 & 0.0295 & 0.0131 & 0.0114 & 0.0145 & 0.3103 & 1.2566 & 0.3777 & 0.0151 & 0.0106 & 0.0089 & 0.0091 & 0.3252 & 1.2705 & 0.3675 \\
\midrule
\multirow{6}{*}{$800{,}000$}
    & \multicolumn{21}{l}{\textbf{LI-ITR}} \\
    & Bias & 0.0445 & 0.0082 & 0.0054 & 0.0095 & 0.0356 & 0.0793 & 0.0927 & 0.0244 & 0.0065 & 0.0035 & 0.0152 & 0.0238 & 0.0504 & 0.1010 & 0.0109 & 0.0028 & 0.0014 & 0.0073 & 0.0091 & 0.0456 & 0.0544 \\
    & SD  & 0.1801 & 0.0189 & 0.0097 & 0.0233 & 0.0760 & 0.3109 & 0.4275 & 0.0621 & 0.0102 & 0.0044 & 0.0447 & 0.0474 & 0.2406 & 0.4713 & 0.0270 & 0.0029 & 0.0012 & 0.0413 & 0.0188 & 0.2715 & 0.3555 \\[1ex]
    & \multicolumn{21}{l}{\textbf{LIME}} \\
    & Bias & 0.0262 & 0.0048 & 0.0296 & 0.0087 & 0.7178 & 1.4695 & 0.5758 & 0.0229 & 0.0149 & 0.0130 & 0.0167 & 0.6309 & 1.4869 & 0.5353 & 0.0076 & 0.0067 & 0.0068 & 0.0075 & 0.7160 & 1.4846 & 0.5230 \\
    & SD  & 0.0171 & 0.0048 & 0.0173 & 0.0082 & 0.3301 & 1.2392 & 0.3778 & 0.0295 & 0.0131 & 0.0115 & 0.0142 & 0.3098 & 1.2567 & 0.3773 & 0.0151 & 0.0104 & 0.0088 & 0.0086 & 0.3248 & 1.2704 & 0.3674 \\
\bottomrule
    \end{tabular}
    \begin{tablenotes}
      \footnotesize
      \item SD is the standard deviation of biases.
    \end{tablenotes}
    \end{threeparttable}
    }}
\end{table}

\begin{table}[ht]
\centering
\caption{Proportion of optimally treated subjects across methods and sample sizes}
\label{tab:pcot_results_cor_quad}
\resizebox{\textwidth}{!}{%
\begin{tabular}{lcccccc}
\hline
& \multicolumn{3}{c}{True model} & \multicolumn{3}{c}{Misspecified local model} \\
\cmidrule(lr){2-4}\cmidrule(lr){5-7}
\textbf{Method} & \textbf{$n=2{,}000$} & \textbf{$n=10{,}000$} & \textbf{$n=40{,}000$} & \textbf{$n=2{,}000$} & \textbf{$n=10{,}000$} & \textbf{$n=40{,}000$} \\
\hline
black-box ITR & 0.989 & 0.995 & 0.999 & 0.986 & 0.992 & 0.993 \\
LI-ITR ($m = 100{,}000$)  & 0.994 & 0.996 & 1.000 & 0.985 & 0.987 & 0.984 \\
LI-ITR ($m = 800{,}000$)  & 0.993 & 0.997 & 0.998 & 0.982 & 0.987 & 0.984 \\

LIME ($m = 100{,}000$)  & 0.878 & 0.888 & 0.881 & 0.950 & 0.949 & 0.948 \\
LIME ($m = 800{,}000$)    & 0.879 & 0.888 & 0.883 & 0.950 & 0.949 & 0.948 \\

OWL (linear) & 0.828 & 0.830 & 0.832 & 0.820 & 0.890 & 0.943 \\
Causal forest & 0.971 & 0.990 & 0.992 & 0.983 & 0.989 & 0.994 \\
Q-learning & 0.836 & 0.832 & 0.837 & 0.937 & 0.936 & 0.937 \\
\hline
\end{tabular}%
}
\end{table}



\begin{table}[ht]
\centering
\caption{Summary statistics of the pre-treatment laboratories and post-treatment AST}\label{baseline_lab}
\begin{tabular}{lccc}
\hline
\textbf{Test} & \textbf{Mean $\pm$ SD} & \textbf{Median (IQR)} & \textbf{Range} \\
\hline
\multicolumn{4}{l}{\textit{Baseline}} \\
\cline{1-1}
ALP (U/L) & 87.5 $\pm$ 60.6 & 77 (63--100) & 25--2,069 \\
ALT (U/L) & 28.1 $\pm$ 23.6 & 22 (15--33) & 2.5--523 \\
AST (U/L) & 24.6 $\pm$ 21.3 & 21 (17--27) & 5--750 \\
Absolute neutrophils ($\times 10^3/\mu$L) & 4.3 $\pm$ 3.2 & 3.5 (2.6--5.0) & 0.1--43.0 \\
WBC ($\times 10^3/\mu$L) & 6.5 $\pm$ 3.7 & 5.7 (4.4--7.5) & 1.3--54.6 \\
BUN (mg/dL) & 13.8 $\pm$ 5.3 & 13 (11--16) & 2--89 \\
Creatinine (mg/dL) & 0.79 $\pm$ 0.25 & 0.80 (0.70--0.80) & 0.40--8.80 \\
Total bilirubin (mg/dL) & 0.46 $\pm$ 0.26 & 0.40 (0.30--0.50) & 0.10--3.80 \\
eGFR (mL/min/1.73m$^2$) & 67.5 $\pm$ 17.2 & 60 (60--61) & 5--123 \\
\hline
\multicolumn{4}{l}{\textit{Post-treatment}} \\
\cline{1-1}
AST (U/L) & 23.6 $\pm$ 14.4 & 20 (17--26) & 5--303 \\
\hline
\end{tabular}
\end{table}

\begin{table}[ht]
\centering
\caption{Distribution of treatment recommendations and estimated policy value}
\label{tab:treatment_dist_policy}
\begin{tabular}{lrrr}
\hline
\textbf{Method} & \textbf{AI count} & \textbf{Tamoxifen count} & \textbf{Value function} \\ \hline

Q-learning      &  85 (19\%)  & 369 (81\%)               & -2.196 \\
OWL (linear)    & 41 (9\%)          & 413 (91\%)               & -2.403 \\
Causal forest   & 0 (0\%)               & 454 (100\%)                & -2.733 \\
LIME           &  30 (7\%)  &   424 (93\%)   & -2.413 \\

black-box ITR &  30 (7\%)   &  424 (93\%)   &  -2.800   \\
LI-ITR &      28 (6\%) &      426 (94\%) & -2.922\\

 \hline
\end{tabular}
\end{table}

\end{document}